\documentclass[a4paper,11pt]{article}
\usepackage[dvips]{graphicx}
\usepackage{amssymb}
\usepackage{amsmath}

\usepackage{cite}

\begin{document}

\title{Superconducting Cosmic Strings and One Dimensional Extended Supersymmetric Algebras}
\author{
V.K. Oikonomou$^{1,}$$^{2}$\,\thanks{voiko@physics.auth.gr}\\
$^1$Department of Mechanical Engineering, Technological Education Institute of Serres\\
62124 Serres, Greece \\
$^2$Department of Theoretical Physics, Aristotle University of Thessaloniki,\\
54124 Thessaloniki, Greece
} \maketitle

\begin{abstract}
In this article we study in detail the supersymmetric structures that underlie the system of fermionic zero modes around a superconducting cosmic string. Particularly, we extend the analysis existing in the literature on the one dimensional $N=2$ supersymmetry and we find multiple $N=2$, $d=1$ supersymmetries. In addition, compact perturbations of the Witten index of the system are performed and we find to which physical situations these perturbations correspond. More importantly, we demonstrate that there exists a much more rich supersymmetric structure underlying the system of fermions with $N_f$ flavors and these are $N$-extended supersymmetric structures with non-trivial topological charges, with ''$N$'' depending on the fermion flavors.
\end{abstract}

\section*{Introduction and Motivation}

Domain walls, cosmic strings and monopoles, are theoretical predictions of various grand unified models beyond the Standard Model \cite{vilenkin,lazaridesvasiko}. Among the three aforementioned topological defects, the most cosmologically acceptable are cosmic strings with a high energy symmetry breaking scale. On the other hand, monopoles and domain walls usually lead to cosmological inconsistencies when observable outcomes, that these defects impose, are taken into account. However, topologically unstable domain walls are phenomenologically acceptable for various reasons. Furthermore, this kind of domain walls are particularly plausible since the presence of a low tension domain wall network in the universe can provide a somehow natural and non-exotic alternative to existing dark energy models related to modified gravity theories (for an important stream of papers on modified gravity theories see \cite{odintsovgravity} and references therein).

A simplified version of the four dimensional super-Poincare algebra is supersymmetric quantum mechanics (abbreviated to SUSY QM hereafter). SUSY QM has become an independent research field, with numerous applications in various research areas (for informative reviews see \cite{reviewsusyqm}). Mathematical aspects of Hilbert spaces corresponding to SUSY QM systems and in addition applications to various quantum mechanical systems were studied in \cite{diffgeomsusyduyalities} and \cite{susyqminquantumsystems,plu1,plu2}. Moreover, extended supersymmetries and their relation to harmonic superspaces or gravity, were presented in \cite{extendedsusy,ivanov}. Applications of SUSY QM to scattering appear in \cite{susyqmscatter} and specific features of supersymmetry breaking were studied in \cite{susybreaking}. The connection between central charge extended SUSY QM and global four dimensional spacetime supersymmetry were studied in \cite{ivanov}.

SUSY QM was originally introduced to model supersymmetry breaking in a four dimensional quantum field theory framework \cite{witten1}. Supersymmetry is a very successful theoretical tool in the beyond the Standard Model physics and also in string theory. The theoretical and phenomenological implications of supersymmetry are so important that rendered it an important ingredient of various research fields \cite{odi1,odi2,odi3,haag,wittentplc,fayet,topologicalcharges,topologicalcharges1,topologicalcharges2}. Up to date there is no experimental verification of supersymmetry and therefore supersymmetry must be broken in our four dimensional world. With regards to supersymmetry breaking, there exist various elegant ways to break supersymmetry and the interested reader may consult \cite{odi1,odi2} and references therein.

The presence of transverse fermionic zero modes around a cosmic string render the string superconducting \cite{supercondstrings}. Superconducting strings play an important role in cosmological frameworks that originate from a grand unified theory model. As is known \cite{supercondstrings}, superconducting current loops coming from GUT superconducting strings at an early epoch, may explain the big voids in the distribution of galaxies and also the formation of galaxies themselves. The connection of the fermionic zero modes around a superconducting cosmic string to $N=2$, $d=1$ supersymmetry was pointed out in reference \cite{oikonomousupercond}. As was shown there, by calculating the Witten index of the underlying supersymmetric fermionic system, suffices to decide whether the string is superconducting or not.

The purpose of this article is to further elaborate on the supersymmetric structures that underlie the fermionic system around superconducting cosmic strings. Particularly, we shall demonstrate that, when we have $N_f$ flavors of fermions, we actually have a total number of $N_f!/(N_f-2)!+N_f$ different $N=2$ SUSY QM algebras underlying the system. We shall see that this gives the Hilbert space of the SUSY QM system an additional symmetry, which actually is product of global $U(1)$ symmetries. Using the $N=2$, $d=1$ SUSY QM algebra we may form higher reducible representations of these algebras. A trivial extension of each $N=2$ SUSY QM algebra by introducing a trivial cental charge is performed too. In addition, we shall study how the Witten index behaves under compact and odd perturbations. This can only be done in the case a background static magnetic field is turned on, with the only non-zero component being $\mathcal{B}_z$. In addition, we provide a criterion to answer if a cosmic string is superconducting, in the case the index $I_q$, frequently used in the bibliography \cite{supercondstrings,oikonomousupercond}, is equal to zero. Finally, we shall demonstrate that the $N=2$ supersymmetries are not the only supersymmetric structure underlying the fermionic system.  Particularly, depending on the fermion flavors, we shall see that there exist extended supersymmetric structures with non-trivial topological charge. These can be $N=4,6,...$ depending on the total number of fermion flavors. With topological charge we mean an operator that is equal to the anticommutator of Fermi charges (supercharges), following the terminology of references \cite{wittentplc,haag,fayet}. As we shall see, these topological charges can be central charges of the supersymmetry algebra when some conditions are satisfied.

This paper is organized as follows: In section 1, we briefly review the theoretical framework of fermions around superconducting strings and also the results from previous work on the relation between the fermionic system with $N=2$, $d=1$ SUSY QM. In section 2, we generalize the results of section 1 in the case we have $N_f$ flavors of fermions and we find multiple $N=2$, $d=1$ supersymmetries in the system. In addition, we demonstrate that the fermionic quantum system has a global $R$-symmetry which is a product of global $U(1)$ symmetries. Moreover, we perform compact odd perturbations on the Witten index and  we also show that we can form higher representations of $N=2$, $d=1$ supersymmetric algebra, by combining the multiple $N=2$ we found previously. Furthermore, we provide an extension of each $N=2$, $d=1$ algebra including a trivial central charge in the algebra. In section 3, we demonstrate that there exists a much more complicated one dimensional supersymmetric structure underlying the fermionic system, which depending on the number of fermion flavors $N_f$, is an $N$-extended one dimensional supersymmetry, with $N=4,6,8,...,2(\frac{N_f!}{(N_f-2)!}+N_f)$. The supersymmetric structures have non-trivial topological charges, which under certain circumstances, can become central charges of the supersymmetric algebra. A small discussion on the issue of topological charges and the implications of these on the supersymmetric structure appears in the end of section 3. The conclusions follow in the end of the article.

\section{Preliminaries-Superconducting Cosmic Strings and N=2 SUSY QM}

In this section we briefly review the general framework of the superconducting cosmic strings models we shall use and the connection of the model with an one dimensional $N=2$ SUSY QM algebra. We follow reference \cite{oikonomousupercond}.  Consider a theory containing $N_f$ left
handed fermion fields $\psi_{\alpha}$ and $N_f$ right-handed
fermions $\chi_{\alpha}$, interacting with the Higgs sector
according to the following Lagrangian,
\begin{equation}\label{lagrangian}
\mathcal{L}=i
\bar{\psi_{\alpha}}\gamma^{\mu}\partial_{\mu}\psi_{\alpha}+i
\bar{\chi_{\alpha}}\gamma^{\mu}\partial_{\mu}\chi_{\alpha}-(\bar{\chi_{\alpha}}M_{\alpha
\beta}\psi_{\beta}+\mathrm{H.c}).
\end{equation}
with $\alpha, \beta =1,...,N_f$. The $N_f\times N_f$ mass-matrix $M_{\alpha \beta}$ contains the scalar fields of the model, that is the Higgs fields, with the interaction couplings. In principle, in a
cosmic string background, the matrix $M$ has explicit dependence on the polar
coordinates $r$ and $\theta$ around the string. Consequently and owing to the
cylindrical symmetry of the string, the theory is effectively two
dimensional and therefore we can work in terms of two component spinors. Hence, the
chiral fermions can be written as,
\begin{equation}\label{psibars}
\psi_{\alpha}=\frac{1}{\sqrt{2}}\left(%
\begin{array}{c}
  \hat{\psi_{\alpha}} \\
  -\hat{\psi_{\alpha}} \\
\end{array}%
\right)
\end{equation}
and also,
\begin{equation}\label{psibars1}
\chi_{\alpha}=\frac{1}{\sqrt{2}}\left(%
\begin{array}{c}
  \hat{\chi_{\alpha}} \\
  -\hat{\chi_{\alpha}} \\
\end{array}%
\right)
\end{equation}
Therefore, the Lagrangian can be written as,
\begin{align}\label{yeaaah}
&\mathcal{L}=i\hat{\psi_{\alpha}}^{\dag}\partial_{0}\psi_{\alpha}-i\hat{\psi_{\alpha}}^{\dag}\sigma^{j}\partial_{j}\psi_{\alpha}
i\hat{\chi_{\alpha}}^{\dag}\partial_{0}\chi_{\alpha}-i\hat{\chi_{\alpha}}^{\dag}\sigma^{j}\partial_{j}\chi_{\alpha}\notag
\\&  - \hat{\chi_{\alpha}^{\dag}}M_{\alpha
\beta}\hat{\psi_{\alpha}}-\hat{\psi_{\alpha}}^{\dag}M_{\alpha
\beta}\hat{\chi_{\alpha}}
\end{align}
The fermionic equations of motion corresponding to the Lagrangian
(\ref{yeaaah}) are,
\begin{align}\label{ref1}
&-\partial_{0}\hat{\psi_{\alpha}}+\sigma^{j}\partial_{j}\hat{\psi_{\alpha}}-iM_{\alpha
\beta}^{\dag}\hat{\chi_{\beta}}=0\\&\notag
-\partial_{0}\hat{\chi_{\alpha}}+\sigma^{j}\partial_{j}\hat{\chi_{\alpha}}-iM_{\alpha
\beta}^{\dag}\hat{\psi_{\beta}}=0
\end{align}
with $\alpha,\beta=1,2,...,N_f$, and $\sigma^j$ the Pauli matrices.
Let,
\begin{equation}\label{ref2}
\hat{\psi_{\alpha}}=f(x_3,t)\left(%
\begin{array}{c}
  \psi_{\alpha}(r,\phi) \\
  0 \\
\end{array}%
\right)
\end{equation}
and additionally,
\begin{equation}\label{ref2222}
\hat{\chi_{\alpha}}=f(x_3,t)\left(%
\begin{array}{c}
  0 \\
  \chi_{\alpha}(r,\phi) \\
\end{array}%
\right)
\end{equation}
Using (\ref{ref2}) and (\ref{ref2222}), the transverse zero-mode equations in the
$x_1{\,}x_2$ plane may be written as,
\begin{align}\label{koryfaia}
&(\partial_{1}+i\partial_{2})\psi_{\alpha}-iM_{\alpha
\beta}^{\dag}\chi_{\beta}=0\\&\notag
(\partial_{1}-i\partial_{2})\chi_{\beta}+iM_{\alpha
\beta}\psi_{\alpha}=0
\end{align}
In addition, one must have,
\begin{equation}\label{yeah}
(\partial_{0}-\partial_{3})f=0
\end{equation}
The last equation means that both $\psi$ and $\chi$ are left
movers (L-movers, see \cite{supercondstrings}). Another possibility is to
have,
\begin{equation}\label{ref2}
\hat{\psi_{\alpha}}=f(x_3,t)\left(%
\begin{array}{c}
  0 \\
  \psi_{\alpha}(r,\phi) \\
\end{array}%
\right)
\end{equation}

\begin{equation}\label{ref2wer}
\hat{\chi_{\alpha}}=f(x_3,t)\left(%
\begin{array}{c}
  \chi_{\alpha}(r,\phi) \\
  0 \\
\end{array}%
\right)
\end{equation}
with corresponding equations of motion,
\begin{align}\label{koryfaiax}
&(\partial_{1}-i\partial_{2})\psi_{\alpha}-iM_{\alpha
\beta}^{\dag}\chi_{\beta}=0\\&\notag
(\partial_{1}+i\partial_{2})\chi_{\beta}+iM_{\alpha
\beta}\psi_{\alpha}=0
\end{align}
and also,
\begin{equation}\label{yeah345}
(\partial_{0}+\partial_{3})f=0
\end{equation}
In this case, both $\psi$ and $\chi$ are right movers (R-movers,
see \cite{supercondstrings}). The focus on this paper is on these zero modes corresponding to solutions of equations (\ref{koryfaia}) and (\ref{koryfaiax}). We define the operator $\mathcal{D}$ to be equal to,
\begin{equation}\label{dmatrix}
\mathcal{D}=\left(%
\begin{array}{cc}
  \partial_{1}+i\partial_{2} & -iM^{\dag} \\
  iM & \partial_{1}-i\partial_{2} \\
\end{array}%
\right)_{2N_f\times 2N_f}
\end{equation}
and in addition,
\begin{equation}\label{dmatrixsdddd}
\mathcal{D}^{\dag}=\left(%
\begin{array}{cc}
  \partial_{1}-i\partial_{2} & -iM^{\dag} \\
  iM & \partial_{1}+i\partial_{2} \\
\end{array}%
\right)_{2N\times 2N}
\end{equation}
acting on,
\begin{equation}\label{wee}
\left(%
\begin{array}{c}
  \psi_{\alpha} \\
  \chi_{\alpha} \\
\end{array}%
\right)
\end{equation}
The solutions of (\ref{koryfaia}) and (\ref{koryfaiax}) correspond to the zero modes of $D$ and also $D^{\dag}$. As was demonstrated in reference \cite{oikonomousupercond}, using the operator $\mathcal{D}$, we can form an $N=2$ SUSY QM algebra with supercharges and Hamiltonian:
\begin{equation}\label{wit2}
Q=\bigg{(}\begin{array}{ccc}
  0 & \mathcal{D} \\
  0 & 0  \\
\end{array}\bigg{)},{\,}{\,}{\,}
Q^{\dag}=\bigg{(}\begin{array}{ccc}
  0 & 0 \\
  \mathcal{D}^{\dag} & 0  \\
\end{array}\bigg{)},{\,}{\,}{\,}
H=\bigg{(}\begin{array}{ccc}
  \mathcal{D}\mathcal{D}^{\dag} & 0 \\
  0 & \mathcal{D}^{\dag}\mathcal{D}  \\
\end{array}\bigg{)}
\end{equation}
which obey, $\{Q,Q^{\dag}\}=H$,
$Q^2=0$, ${Q^{\dag}}^2=0$, $\{Q,W\}=0$, $W^2=I$ and $[W,H]=0$. This result has important phenomenological consequences which we shall not discuss here. For details see \cite{oikonomousupercond}. Our interest in this paper is mainly on the extra hidden extended SUSY QM structure of the fermion-Higgs system around the cosmic string which we will start to reveal in the next section.

\section{A Model with $N_f$ flavors of Fermions-Many $N=2$ Supersymmetries and Implications}

The starting point of our analysis shall be equations (\ref{koryfaia}) and (\ref{koryfaiax}). We shall assume that the system consists of $N_f$ different fermion flavors and also that the mass matrix has the general form: 
\begin{equation}\label{indexofd}
M_{\alpha \beta}(r,\phi)=S_{\alpha \beta}(r)e^{iq_{\alpha
\beta}\phi}
\end{equation}
The integers $q_{\alpha \beta}$ are directly related to the charges of the
fields with respect to the group generator $Q$ which corresponds
to the cosmic string \cite{supercondstrings}. Let $\bar{q}_{\alpha}$ and
$q_{\beta}$ the charges of the fermion fields
$\chi_{\alpha}^{\dag}$ and $\psi_{\beta}$, then \cite{supercondstrings,oikonomousupercond}:
\begin{equation}\label{tetanus}
q_{\alpha \beta}=\bar{q}_{\alpha}-q_{\beta}
\end{equation}
In any case, the mass matrix $M_{\alpha \beta}$ depends on the coordinates $(r,\phi)$, unless otherwise stated. What was not pointed out in \cite{oikonomousupercond} is that there exist $N_s$ different $N=2$, $d=1$ supersymmetries and not just one, with $N_s$ equal to:
\begin{equation}\label{ndheestuff}
N_s=\frac{N_f!}{(N_f-2)!}+N_f
\end{equation}
Indeed, consider equations (\ref{koryfaia}) which relate $\chi_{\beta}$ and $\psi_{\alpha}$. If we fix $\alpha$ and $\beta$ with $\alpha,\beta =1,2,...,N_f$, we can form the $2\times 2$ operator $\mathcal{D}_{\alpha \beta}$, which is equal to:
\begin{equation}\label{dmatrixnew}
\mathcal{D}_{\alpha \beta}=\left(%
\begin{array}{cc}
  \partial_{1}+i\partial_{2} & -iM^{\dag}_{\alpha \beta} \\
  iM_{\alpha \beta} & \partial_{1}-i\partial_{2} \\
\end{array}%
\right)_{2\times 2}
\end{equation}
and in addition,
\begin{equation}\label{dmatrixnewss}
\mathcal{D}^{\dag}_{\alpha \beta}=\left(%
\begin{array}{cc}
  \partial_{1}-i\partial_{2} & -iM^{\dag}_{\alpha \beta} \\
  iM_{\alpha \beta} & \partial_{1}+i\partial_{2} \\
\end{array}%
\right)_{2\times 2}
\end{equation}
acting on,
\begin{equation}\label{wedddde}
\left(%
\begin{array}{c}
  \psi_{\alpha} \\
  \chi_{\beta} \\
\end{array}%
\right)
\end{equation}
Following the line of research of the previous section, using the operator $\mathcal{D}$ we can form an $N=2$ SUSY QM algebra with supercharges and Hamiltonian:
\begin{equation}\label{wit2jdnhd}
\mathcal{Q}_{\alpha \beta}=\bigg{(}\begin{array}{ccc}
  0 & D_{\alpha \beta} \\
  0 & 0  \\
\end{array}\bigg{)},{\,}{\,}{\,}
\mathcal{Q}^{\dag}_{\alpha \beta}=\bigg{(}\begin{array}{ccc}
  0 & 0 \\
  D^{\dag}_{\alpha \beta} & 0  \\
\end{array}\bigg{)},{\,}{\,}{\,}
\mathcal{H}_{\alpha \beta}=\bigg{(}\begin{array}{ccc}
  \mathcal{D}_{\alpha \beta}\mathcal{D}_{\alpha \beta}^{\dag} & 0 \\
  0 & \mathcal{D}_{\alpha \beta}^{\dag}\mathcal{D}_{\alpha \beta}  \\
\end{array}\bigg{)}
\end{equation}
which obey, 
\begin{equation}\label{dnegexhhcch}
\{\mathcal{Q}_{\alpha \beta},\mathcal{Q}_{\alpha \beta}^{\dag}\}=\mathcal{H}_{\alpha \beta},{\,}{\,}{\,}\mathcal{Q}_{\alpha \beta}^2=0,{\,}{\,}{\,}{\mathcal{Q}_{\alpha \beta}^{\dag}}^2=0,{\,}{\,}{\,} \{\mathcal{Q}_{\alpha \beta},W\}=0,{\,}{\,}{\,}W^2=I,{\,}{\,}{\,}[W,\mathcal{H}_{\alpha \beta}]=0
\end{equation}
In the present case, the Witten operator $W$ is equal to the $2\times 2$ matrix:
\begin{equation}\label{s7345}
W=\bigg{(}\begin{array}{ccc}
  1 & 0 \\
  0 & -1  \\
\end{array}\bigg{)}
\end{equation}
Notice that we can form $N_s=N_f!/(N_f-2)!+N_f$ such algebras, since we have $N_s$ different complex supercharges. Indeed, for the case $N_f=2$ (four fermions totally), the supercharges are the following:
\begin{equation}\label{nfsupch}
\mathcal{Q}_{11},{\,}{\,}{\,}\mathcal{Q}_{12},{\,}{\,}{\,}\mathcal{Q}_{21},{\,}{\,}{\,}\mathcal{Q}_{22}
\end{equation}
while for $N_f=4$ (eight fermions totally),
\begin{align}\label{nfsu111pch}
&\mathcal{Q}_{11},{\,}{\,}{\,}\mathcal{Q}_{12},{\,}{\,}{\,}\mathcal{Q}_{13},{\,}{\,}{\,}\mathcal{Q}_{14}\\ \notag &
\mathcal{Q}_{21},{\,}{\,}{\,}\mathcal{Q}_{22},{\,}{\,}{\,}\mathcal{Q}_{23},{\,}{\,}{\,}\mathcal{Q}_{24}\\ \notag &
\mathcal{Q}_{31},{\,}{\,}{\,}\mathcal{Q}_{32},{\,}{\,}{\,}\mathcal{Q}_{33},{\,}{\,}{\,}\mathcal{Q}_{34} \\ \notag &
\mathcal{Q}_{41},{\,}{\,}{\,}\mathcal{Q}_{42},{\,}{\,}{\,}\mathcal{Q}_{43},{\,}{\,}{\,}\mathcal{Q}_{44}
\end{align}
and so on. Now this supersymmetric structure has some important implications for the fermionic Hilbert space formed by the zero modes of the operators $\mathcal{D}_{\alpha \beta}$, which were not pointed out in \cite{oikonomousupercond}. Particularly, as we demonstrate in the next subsection, each $N=2$, $d=1$ SUSY QM algebra with supercharge $\mathcal{Q}_{\alpha \beta}$ provides the Hilbert space of the fermions $(\psi_{\alpha},\chi_{\beta})$ with a global $U(1)$ symmetry.

\subsection{A Global R-symmetry of the Quantum Hilbert Space}

The existence of an $N=2$, $d=1$ SUSY QM algebra on the fermionic space formed by the fermions $(\psi_{\alpha},\chi_{\beta})$, has a direct implication on the Hilbert space of these quantum states. Particularly, the quantum algebra is invariant under a global $U(1)$ symmetry. In order to see this, we perform the following transformation on the supercharges ${\mathcal{Q}}_{\alpha \beta}$ and ${\mathcal{Q}}^{\dag}_{\alpha \beta}$ (we fix the values of ${\alpha, \beta}$):
\begin{align}\label{transformationu1}
& {\mathcal{Q}}_{\alpha \beta}^{'}=e^{-ia_{\alpha \beta}}{\mathcal{Q}}_{\alpha \beta}, {\,}{\,}{\,}{\,}{\,}{\,}{\,}{\,}
{\,}{\,}{\mathcal{Q}}^{'\dag}_{\alpha \beta}=e^{ia_{\alpha \beta}}{\mathcal{Q}}^{\dag}_{\alpha \beta}
.\end{align}
The quantum Hamiltonian of the system, namely $\mathcal{H}_{\alpha \beta}$, is invariant under this global transformation, but the quantum states of the system are transformed as follows:
\begin{equation}\label{depeche}
\psi^{'+}_{\alpha \beta}=e^{-i\beta^{+}_{\alpha \beta}}\psi^{+}_{\alpha \beta},
{\,}{\,}{\,}{\,}{\,}{\,}{\,}{\,}
{\,}{\,}\psi^{'-}_{\alpha \beta}=e^{-i\beta^{-}_{\alpha \beta}}\psi^{-}_{\alpha \beta}
.\end{equation}
Of course, the parameters $\beta^{+}_{\alpha \beta}$ and $\beta^{-}_{\alpha \beta}$ are global
parameters that are connected to $a_{\alpha \beta}$ as $a_{\alpha \beta}=\beta^{+}_{\alpha \beta}-\beta^{-}_{\alpha \beta}$.
In relation (\ref{depeche}), the quantum states $\psi^{+}_{\alpha \beta}$ and $\psi^{-}_{\alpha \beta}$ are the Witten parity even and odd quantum states, which are related to the fermionic fields $(\psi_{\alpha},\chi_{\beta})$ (see \cite{oikonomousupercond} for details).  Taking into account that for $N_f$ fermionic flavors, we have $N_s=N_f!/(N_f-2)!+N_f$ such algebras, we conclude that the total symmetry $U_{tot}$ of the system is the product of $N_f!/(N_f-2)!+N_f$ different global $U(1)$ symmetries, that is:
\begin{equation}\label{totsymm}
U_{tot}=\underbrace{U(1)_{11}\times U_{12}\times ...U_{N_fN_f}}_{N_f!/(N_f-2)!+N_f{\,}\mathrm{times}}
\end{equation}

\subsection{Witten Index, $N=2$ SUSY QM and Static Background Electromagnetic Fields}

The solutions of equations (\ref{koryfaia}) and (\ref{koryfaiax}) are the
zero modes of the operators $\mathcal{D}$ and $\mathcal{D}^{\dag}$. The Fredholm index $I$ of
the operator $\mathcal{D}$, is defined to be:
\begin{equation}\label{indexd}
\mathrm{ind}\mathcal{D}=\mathrm{I}_q=\mathrm{dim{\,}ker}(\mathcal{D}^{\dag})-\mathrm{dim{\,}ker}(\mathcal{D})
\end{equation}
which is the number of zero modes of $\mathcal{D}$ (L-movers) minus the
number of zero modes of $\mathcal{D}^{\dag}$ (R-movers). In terms of the transverse zero modes around the string \cite{oikonomousupercond}, this index is equal to the number of the right movers $R$ minus the number of the left movers
$L$. In references \cite{supercondstrings} it is
proved that $I_q=\sum_{\alpha =1}^{n}q_{\alpha \alpha}$. Therefore, the
Fredholm index of $\mathcal{D}$ is related to the charges of the fermions, which are directly related to
the string gauge group. In reference \cite{oikonomousupercond} it was established that the Witten index of the $N=2$ supersymmetric quantum mechanics system is related to the index $I_q$ of the charges that
the fermions have. Indeed, $I_q=-\Delta$, owing to the fact that,
\begin{equation}\label{ker}
I_q=\mathrm{dim}{\,}\mathrm{ker}
\mathcal{D}^{\dag}-\mathrm{dim}{\,}\mathrm{ker} \mathcal{D}=
\mathrm{dim}{\,}\mathrm{ker}\mathcal{D}\mathcal{D}^{\dag}-\mathrm{dim}{\,}\mathrm{ker}\mathcal{D}^{\dag}\mathcal{D}=-\mathrm{ind}\mathcal{D}=-\Delta=n_--n_+
\end{equation}
This result is phenomenologically valuable since the underlying supersymmetric algebra is directly
related to the phenomenology of the model on which the
superconducting string is built upon. Particularly, we may answer the question if there exists string superconductivity by examining the Witten index of the underlying $N=2$ supersymmetric algebra. Actually, it is known that when
$I_q\neq 0$, string superconductivity is guaranteed \cite{supercondstrings}. According
to relation (\ref{ker}), string superconductivity occurs always when the
Witten index $\Delta$ is non-zero, which corresponds to unbroken $N=2$ SUSY QM. Therefore, it suffices to have unbroken SUSY QM in the model under examination in order to guarantee string superconductivity. Recall that the transverse zero modes of the fermions are the zero modes of operators (\ref{dmatrix}) ($L$-movers) and (\ref{dmatrixsdddd}) ($R$-movers).

In this section we further study the connection of the Witten index with the $I_q$ index defined in references \cite{supercondstrings}. Particularly, we shall address two questions having to do with the Witten index and with the index $I_q$. These questions were not answered in \cite{oikonomousupercond}. Firstly, if there is string superconductivity when $I_q=0$. It is known that when $I_q=0$, it is not sure whether superconducting strings exist or
not \cite{supercondstrings}. The second question has to do with perturbations of the Witten index when static background electric and magnetic fields are taken into account. We shall address these questions in the following subsections.

\subsubsection{String Superconductivity when $I_q=0$}

The operators we are dealing with are by definition Fredholm since the have a finite kernel. Let us work with the operators (\ref{dmatrix}) and (\ref{dmatrixsdddd}), but the same applies for the operators (\ref{dmatrixnew}) and (\ref{dmatrixnewss}). Let $n_-,n_+$ be the number of zero modes of the operators $\mathcal{D}^{\dag}$ and $\mathcal{D}$ respectively. Since the operators are Fredholm (which implies a finite number of $n_{\pm}$ zero modes) the quantity,
\begin{equation}\label{phil}
\Delta =n_{-}-n_{+}
\end{equation}
is actually the Witten index. Whenever the Witten index is non-zero
integer, supersymmetry is unbroken. If the Witten index is
zero, the we have two possibilities. If $n_{+}=n_{-}=0$ then supersymmetry is broken but if $n_{+}= n_{-}\neq 0$, supersymmetry is unbroken. Thus we may conclude that since $I_q=-\Delta$, when $I_q=0$ and $n_{+}= n_{-}\neq 0$, string superconductivity is guaranteed, since we have zero modes of the operators $\mathcal{D}^{\dag}$ and $\mathcal{D}$, but they happen to be equal in number. The same result applies in the case in which we examine the operators (\ref{dmatrixnew}) and (\ref{dmatrixnewss}). Notice that in this case, the study is very much simplified since string superconductivity may be studied in a sub-sector of the whole system, namely that of the fermions $(\chi_{\beta},\psi_{\alpha})$ with fixed $(\alpha,\beta)$.

\subsubsection{Compact Perturbations of the Witten Index}

One more question we shall address in this section is whether the Witten index remains invariant when we turn on static background electromagnetic fields. The study of superconducting strings in the presence of electromagnetic fields can be found in \cite{witten2}. A solid criterion in order to investigate this problem is a theorem that deals with compact perturbations of the index of Fredholm operators. We shall focus on the subsystem that consists of two fermions, which corresponds to equations:
\begin{align}\label{koryfaiaxwe}
&(\partial_{1}-i\partial_{2})\psi_{\alpha}-iM_{\alpha
\beta}^{\dag}\chi_{\beta}=0\\&\notag
(\partial_{1}+i\partial_{2})\chi_{\beta}+iM_{\alpha
\beta}\psi_{\alpha}=0
\end{align}
with fixed $\alpha ,\beta$. Notice that this set of equations gives the zero modes of the operator (\ref{dmatrixnew}), which is Fredholm. Let $\mathcal{C}$ an odd compact symmetric matrix \cite{thaller} of the form:
\begin{equation}\label{susyqmrnmassive}
\mathcal{C}=\left(%
\begin{array}{cc}
 0 & \mathcal{C}_1
 \\ \mathcal{C}_2 & 0\\
\end{array}%
\right)
\end{equation}
Consider the operator $\mathcal{D}_{p}=\mathcal{D}_{\alpha \beta}+\mathcal{C}$, which is actually a compact perturbation of the operator $\mathcal{D}_{\alpha \beta}$. Since the operator $\mathrm{tr}\mathcal{W}e^{-t(\mathcal{D}_1+\mathcal{C})^2}$ is trace class (owing to the fact that compact perturbations of Fredholm operators are also Fredholm operators and therefore trace-class), the following theorem holds true (see \cite{thaller} page 168, Theorem 5.28),
\begin{equation}\label{indperturbatrn1}
\mathrm{ind}\mathcal{D}_p=\mathrm{ind}(\mathcal{D}_{\alpha \beta}+\mathcal{C})=\mathrm{ind}\mathcal{D}_{\alpha \beta}
,\end{equation}
with $\mathcal{C}$ the symmetric odd operator of the form (\ref{susyqmrnmassive}). Hence, owing to the theorem (\ref{indperturbatrn1}) and due to relation (\ref{ker}), the Witten index of the SUSY QM system corresponding to operator $\mathcal{D}_{\alpha \beta}$, is invariant under compact perturbations of the operator $\mathcal{D}_{\alpha \beta}$. The question now is which kind of static background electromagnetic fields generate compact perturbations. As we shall see, only the case corresponding to turning on a background fast decaying of constant static magnetic field leaves the Witten index invariant.

It is not so difficult to show that the addition background electric fields, the operator $\mathcal{D}_{\alpha \beta}$ is modified as follows:
\begin{equation}\label{noodd}
\mathcal{D}_{\alpha \beta}+\mathcal{F}
\end{equation}
with $\mathcal{F}$ an even matrix. Owing to the fact that the matrix $\mathcal{F}$ is even, the theorem (\ref{indperturbatrn1}) does not apply and hence, the Witten index does not remain invariant. In the case we turn on a magnetic field of the form $\mathcal{B}=(\mathcal{B}_x,\mathcal{B}_y,\mathcal{B}_z)$, the first two components modify equations (\ref{koryfaiaxwe}) in such a way so that operator $\mathcal{D}_{\alpha \beta}$ is modified in the same way as in equation (\ref{noodd}), and hence, the operator $\mathcal{D}_{\alpha \beta}$ is perturbed by the addition of an even compact operator. Nevertheless, only the last component of $\mathcal{B}$, generates compact and odd perturbations of operator $\mathcal{D}_{\alpha \beta}$. We demonstrate this case in detail. Firstly, let us note that the use of fast decaying magnetic fields is necessary in order the operator perturbations are compact. Now consider an electromagnetic field of the form $\mathcal{A}^{\mu}=(0,0,0,\mathcal{B}_z)$, with $\mathcal{B}_z$ a constant or fast decaying magnetic field (constant or fast decaying in order the operators are compact, as we already pointed out). Including the background magnetic field interaction with the fermions, the equations (\ref{koryfaiaxwe}) can be written as follows:
\begin{align}\label{koryfaiaxwe12}
&(\partial_{1}-i\partial_{2})\psi_{\alpha}-(iM_{\alpha
\beta}^{\dag}-\mathcal{B}_z)\chi_{\beta}=0\\&\notag
(\partial_{1}+i\partial_{2})\chi_{\beta}+(iM_{\alpha
\beta}-\mathcal{B}_z)\psi_{\alpha}=0
\end{align}
Consequently, the operator $\mathcal{D}_{\alpha \beta}$ acquires an additional contribution and can be written as follows:
\begin{equation}\label{chaniaaddop}
\mathcal{D}_{\alpha \beta}'=\mathcal{D}_{\alpha \beta}+\mathcal{C}_z,
\end{equation}
with $\mathcal{C}_z$ being equal to:
\begin{equation}\label{susyqmrnmafgreh}
\mathcal{C}_z=\left(%
\begin{array}{cc}
 0 & \mathcal{B}_z
 \\ -\mathcal{B}_z & 0\\
\end{array}%
\right)
\end{equation}
The operator $\mathcal{C}_z$ is odd and in order theorem (\ref{indperturbatrn1}) holds true, we must additionally require that the operator $\mathcal{B}_z$ is compact. As we already pointed out, this can be true when $\mathcal{B}_z$ is a finite constant, or if it decays fast. The functional dependence of $\mathcal{B}_z$ is of no mathematical importance, but the physics underlying the choice of $\mathcal{B}_z$ can be potentially interesting, owing to the fact that it has to do with a magnetic field near, or on a cosmic string, along the $z$-direction. We have no interest in the cosmological implications of this, but we are interested in the mathematical implications of compact perturbations of the system. Hence, if the operator $\mathcal{B}_z$ is compact, according to theorem (\ref{indperturbatrn1}) the Witten index of the system is invariant, that is:
\begin{equation}\label{fhjhggfgy60seconds}
\mathrm{ind}\mathcal{D}_{\alpha \beta}'=\mathrm{ind}(\mathcal{D}_{\alpha \beta}+\mathcal{C}_z)=\mathrm{ind}\mathcal{D}_{\alpha \beta}
\end{equation} 
This means that the net number of the transverse fermionic zero modes of the system along the $z$-direction of the string, remains unaltered even if we add a magnetic field with the only non-zero component being parallel to the $z$-direction of the cosmic string. So the presence of a magnetic field with the aforementioned characteristics does not alter the net number of zero modes of the fermionic system.

\subsection{Extended Supersymmetric-Higher Representation Algebras}

Up to this point we demonstrated that the fermionic zero modes of the fermion fields $(\chi_{\alpha},\psi_{\beta})$ around a superconducting cosmic string can be related to a large number of $N=2$ SUSY QM algebras, depending on the fermion flavors $N_f$. Particularly, if each of the fermion fields $(\chi_{\alpha},\psi_{\beta})$ have $N_f$ flavors, then the total number of different $N=2$ SUSY QM algebras is $N_f!/(N_f-2)!+N_f$. The question is whether these SUSY QM algebras can combine to form extended supersymmetries or higher dimensional SUSY QM representations. The answer to both question is yes and we address the latter question in this section.

\subsubsection{Higher Reducible Representations I}

\noindent Consider two of the $N_f!/(N_f-2)!+N_f$ different complex supercharges $(\mathcal{Q}_{\alpha \beta},\mathcal{Q}_{\alpha ' \beta '})$ and their corresponding differential operators $({\mathcal{D}}_{\alpha \beta},{\mathcal{D}}_{\alpha ' \beta '})$. The values of $\alpha , \beta ,\alpha ' \beta ' $ are in general different but some of these can be equal (for example in the case the supercharges $\mathcal{Q}_{11},\mathcal{Q}_{12}$ are examined). The two $N=2$ supersymmetries corresponding to these supercharges can form a higher reducible representation of a single $N=2$, $d=1$ supersymmetry. We denote ${\mathcal{Q}}_{DU}$ and  ${\mathcal{Q}}_{DU}^{\dag}$ the supercharges of this higher order $N=2$ SUSY QM algebra, which are equal to:
\begin{equation}\label{connectirtyrtons}
{\mathcal{Q}}_{DU}= \left ( \begin{array}{cccc}
  0 & 0 & 0 & 0 \\
  {\mathcal{D}}_{\alpha  \beta } & 0 & 0 & 0 \\
0 & 0 & 0 & 0 \\
0 & 0 & {\mathcal{D}}_{\alpha ' \beta '}^{\dag} & 0  \\
\end{array} \right),{\,}{\,}{\,}{\,}{\mathcal{Q}}_{DU}^{\dag}= \left ( \begin{array}{cccc}
  0 &  {\mathcal{D}}_{\alpha  \beta }^{\dag} & 0 & 0 \\
  0 & 0 & 0 & 0 \\
0 & 0 & 0 & {\mathcal{D}}_{\alpha ' \beta '} \\
0 & 0 & 0 & 0  \\
\end{array} \right)
.\end{equation}
Moreover, the Hamiltonian of the combined quantum system, which we denote $\mathcal{H}_{DU}$, reads,
\begin{equation}\label{connections1dtr}
H_{DU}= \left ( \begin{array}{cccc}
  {\mathcal{D}}_{\alpha ' \beta '}^{\dag}{\mathcal{D}}_{\alpha ' \beta '} & 0 & 0 & 0 \\
  0 & {\mathcal{D}}_{\alpha ' \beta '}{\mathcal{D}}_{\alpha ' \beta '}^{\dag} & 0 & 0 \\
0 & 0 & {\mathcal{D}}_{\alpha \beta }{\mathcal{D}}_{\alpha \beta }^{\dag} & 0 \\
0 & 0 & 0 & {\mathcal{D}}_{\alpha \beta }^{\dag}{\mathcal{D}}_{\alpha \beta }  \\
\end{array} \right)
.\end{equation}
The above operators (\ref{connectirtyrtons}) and (\ref{connections1dtr}), satisfy the $N=2$, $d=1$ SUSY QM algebra:
\begin{equation}\label{mousikisimagne}
\{ {\mathcal{Q}}_{DU},{\mathcal{Q}}_{DU}^{\dag}\}=H_{DU},{\,}{\,}{\mathcal{Q}}_{DU}^2=0,{\,}{\,}{{\mathcal{Q}}_{DU}^{\dag}}^2=0,{\,}{\,}\{{\mathcal{Q}}_{DU},\mathcal{W}_{DU}\}=0,{\,}{\,}\mathcal{W}_{DU}^2=I,{\,}{\,}[\mathcal{W}_{DU},H_{DU}]=0
.\end{equation}
In the present case, the Witten parity operator is:
\begin{equation}\label{wparityopera}
\mathcal{W}_{DU}= \left ( \begin{array}{cccc}
  1 & 0 & 0 & 0 \\
  0 & -1 & 0 & 0 \\
0 & 0 & 1 & 0 \\
0 & 0 & 0 & -1  \\
\end{array} \right)
.\end{equation}
Additionally, we can form other equivalent higher dimensional representations of the combined $N=2$, $d=1$ algebra, by employing the following set of transformations:
\begin{equation}\label{setof transformations}
\mathrm{Set}{\,}{\,}{\,}A:{\,}
\begin{array}{c}
 {\mathcal{D}}_{\alpha ' \beta '}\rightarrow {\mathcal{D}}_{\alpha ' \beta '}^{\dag} \\
  {\mathcal{D}}_{\alpha \beta }^{\dag}\rightarrow {\mathcal{D}}_{\alpha \beta } \\
\end{array},{\,}{\,}{\,}\mathrm{Set}{\,}{\,}{\,}B:{\,}
\begin{array}{c}
 {\mathcal{D}}_{\alpha ' \beta '}\rightarrow {\mathcal{D}}_{\alpha  \beta }^{\dag} \\
  {\mathcal{D}}_{\alpha \beta }^{\dag}\rightarrow {\mathcal{D}}_{\alpha ' \beta '} \\
\end{array},{\,}{\,}{\,}\mathrm{Set}{\,}{\,}{\,}C:{\,}
\begin{array}{c}
 {\mathcal{D}}_{\alpha ' \beta '}\rightarrow {\mathcal{D}}_{\alpha \beta } \\
  {\mathcal{D}}_{\alpha \beta }^{\dag}\rightarrow {\mathcal{D}}_{\alpha ' \beta '}^{\dag} \\
\end{array}
.\end{equation}
Furthermore, another equivalent to the one of relation (\ref{connectirtyrtons}) higher order reducible representation of the $N=2$ SUSY QM algebra, is given by: 
\begin{equation}\label{connectirtyfhfghrtons}
{\mathcal{Q}}_{DU}= \left ( \begin{array}{cccc}
  0 & 0 & 0 & 0 \\
  0 & 0 & 0 & 0 \\
{\mathcal{D}}_{\alpha ' \beta '} & 0 & 0 & 0 \\
0 & {\mathcal{D}}_{\alpha \beta }^{\dag} & 0 & 0  \\
\end{array} \right),{\,}{\,}{\,}{\,}{\mathcal{Q}}_{DU}^{\dag}= \left ( \begin{array}{cccc}
  0 & 0 & {\mathcal{D}}_{\alpha ' \beta '}^{\dag} & 0 \\
   & 0 & 0 & {\mathcal{D}}_{\alpha \beta } \\
0 & 0 & 0 & 0 \\
0 & 0 & 0 & 0  \\
\end{array} \right)
.\end{equation}
with the Hamiltonian in this case being equal to,
\begin{equation}\label{connectihgghdhtons1dtr}
H_{DU}= \left ( \begin{array}{cccc}
  {\mathcal{D}}_{\alpha ' \beta '}^{\dag}{\mathcal{D}}_{\alpha ' \beta '} & 0 & 0 & 0 \\
  0 & {\mathcal{D}}_{\alpha \beta }^{\dag}{\mathcal{D}}_{\alpha \beta } & 0 & 0 \\
0 & 0 & {\mathcal{D}}_{\alpha  \beta }{\mathcal{D}}_{\alpha  \beta }^{\dag} & 0 \\
0 & 0 & 0 & {\mathcal{D}}_{\alpha ' \beta '}{\mathcal{D}}_{\alpha ' \beta '}^{\dag}  \\
\end{array} \right)
.\end{equation}
Obviously, similar considerations can be done for any pair of the $N_f!/(N_f-2)!+N_f$ available different complex supercharges.

\subsubsection{Higher Reducible Representations II}

In the case we just demonstrated, in order to form higher order reducible representations, we made use of the operators $\mathcal{D}_{\alpha \beta}$ and $\mathcal{D}_{\alpha ' \beta '}$, with fixed $(\alpha ,\beta$). In this subsection we shall make use of the corresponding supercharges directly and we shall form higher order representations. Using the supercharges $(\mathcal{Q}_{\alpha \beta },\mathcal{Q}_{\alpha ' \beta '})$ we can form the following higher order supercharges:
\begin{equation}\label{s7supcghagemixdesec}
\mathcal{Q}_{DU}=\bigg{(}\begin{array}{ccc}
  0 & \mathcal{Q}_{\alpha ' \beta '} \\
  \mathcal{Q}_{\alpha \beta } & 0  \\
\end{array}\bigg{)},{\,}{\,}{\,}\mathcal{Q}_{DU}^{\dag}=\bigg{(}\begin{array}{ccc}
  0 &  \mathcal{Q}_{\alpha \beta }^{\dag} \\
 \mathcal{Q}_{\alpha ' \beta '}^{\dag} & 0  \\
\end{array}\bigg{)}
\end{equation}
The corresponding quantum Hamiltonian reads:
\begin{equation}\label{s11fgghhfdesec}
\mathcal{H}_{DU}=\bigg{(}\begin{array}{ccc}
 \mathcal{Q}_{\alpha ' \beta '}\mathcal{Q}_{\alpha ' \beta '}^{\dag}+ \mathcal{Q}_{\alpha \beta }^{\dag}\mathcal{Q}_{\alpha \beta }& 0 \\
  0 & \mathcal{Q}_{\alpha ' \beta '}^{\dag}\mathcal{Q}_{\alpha ' \beta '}+ \mathcal{Q}_{\alpha \beta }\mathcal{Q}_{\alpha \beta }^{\dag}  \\
\end{array}\bigg{)}
\end{equation}
Of course, the new supercharges and the Hamiltonian satisfy the $N=2$, $d=1$ algebra (\ref{mousikisimagne}).

\subsection{Localized Fermions and SUSY QM-Trivial Central Charge Case}

So far in this article, we saw that the underlying supersymmetric structures have no central charge. However, as we will show in this section, we can enrich each $N=2$ SUSY QM algebra with a real trivial central charge (we consider the ''central charge'' being an operator equal to the anticommutator of two Fermi charges, following \cite{haag} and also commuting with every element of the algebra). Let us take one supercharge from the total number of $N_f!/(N_f-2)!+N_f$ different complex supercharges, namely $\mathcal{Q}_{\alpha \beta }$. We can extend the algebra of relation (\ref{dnegexhhcch}) to include a real central charge, which we denote $Z$, in the following way:
\begin{equation}\label{s7supcghagemixcentch}
\mathcal{Q}_{Z}=\bigg{(}\begin{array}{ccc}
  -\eta & 0 \\
  \mathcal{D}_{\alpha \beta } & \eta  \\
\end{array}\bigg{)},{\,}{\,}{\,}\mathcal{Q}_{Z}=\bigg{(}\begin{array}{ccc}
  -\eta &  \mathcal{D}_{\alpha \beta }^{\dag} \\
 0 & \eta \\
\end{array}\bigg{)}
\end{equation}
with ''$\eta$'' an arbitrary $2\times 2$ real matrix. The Hamiltonian of the system in this case reads:
\begin{equation}\label{s11fgghhfcentcharge}
\mathcal{H}_{Z}=\bigg{(}\begin{array}{ccc}
 \mathcal{D}_{\alpha \beta }\mathcal{D}_{\alpha \beta }^{\dag}+2\eta^2 & 0 \\
  0 & \mathcal{D}_{\alpha \beta }^{\dag}\mathcal{D}_{\alpha \beta }+2\eta^2  \\
\end{array}\bigg{)}
\end{equation}
The real central charge of the centrally extended $N=2$ SUSY QM algebra is equal to:
\begin{equation}\label{centrextendcc}
Z=\bigg{(}\begin{array}{ccc}
  2\eta^2 & 0 \\
  0 & 2\eta^2  \\
\end{array}\bigg{)}.
\end{equation}
Now, the new form of the trivial central charge extended $N=2$ SUSY QM algebra is governed by the following commutation relations:
\begin{equation}\label{relationsforsusycc}
\{\mathcal{Q}_{Z},\mathcal{Q}^{\dag}_{Z}\}=\mathcal{H}_{Z}{\,}{\,},\{\mathcal{Q}_{Z},\mathcal{Q}_{Z}\}=Z,{\,}{\,}\{\mathcal{Q}^{\dag}_{Z},\mathcal{Q}^{\dag}_{Z}\}=Z,{\,}{\,}{\,}[\mathcal{H}_{Z},\mathcal{Q}_{Z}]=[\mathcal{H}_{Z},\mathcal{Q}_{Z}^{\dag}]=0
\end{equation}
The question is what new features this central charge brings to the Hilbert space of the SUSY QM algebra (\ref{relationsforsusycc}). The answer is that the new supercharges no longer map the parity-even to parity-odd states, in reference to the non-zero modes. In the case the matrix $2\eta^{2}$ is an odd compact matrix, then the index of the operator $\mathcal{D}_{\alpha \beta }$ is invariant under the compact perturbation generated by the matrix $2\eta^2$. We can easily see this by exploiting theorem (\ref{indperturbatrn1}). Owing to the fact that the matrix $\eta^2$ is compact and odd, theorem (\ref{indperturbatrn1}) holds true and hence, the following relations hold true: 
\begin{align}\label{genkerrelationscompnew}
&-\Delta = \mathrm{dim}{\,}\mathrm{ker}\mathcal{D}_{\alpha \beta }^{\dag}\mathcal{D}_{\alpha \beta }-\mathrm{dim}{\,}\mathrm{ker}\mathcal{D}_{\alpha \beta }\mathcal{D}_{\alpha \beta }^{\dag}=
\\ \notag & \mathrm{dim}{\,}\mathrm{ker}(\mathcal{D}_{\alpha \beta }^{\dag}\mathcal{D}_{\alpha \beta }+2\eta^2)-\mathrm{dim}{\,}\mathrm{ker}(\mathcal{D}_{\alpha \beta }\mathcal{D}_{\alpha \beta }^{\dag}+2\eta^2)
\end{align}
Therefore, the Witten index of the non-central charge supersymmetric quantum mechanical system is invariant under the central charge extension of the system. This holds true for a real central charge and in addition, with the relations (\ref{centrextendcc}), (\ref{etasquareoddcccc}) and (\ref{etasquareoddcccceta}) simultaneously holding true. This result is valid since both the operators $\mathcal{D}_{\alpha \beta }\mathcal{D}_{\alpha \beta }^{\dag}$ and $\mathcal{D}_{\alpha \beta }^{\dag}\mathcal{D}_{\alpha \beta }$ are trace-class (product of Fredholm--trace-class operators) and therefore the theorem (\ref{indperturbatrn1}) applies to each of them.

Before closing, let us examine how the constraints ''compact'' and ''odd'', affect the matrix $2\eta^2$. Compact means that the matrix $\eta$ must contain elements which are finite numbers or functions. Odd means that $\eta^2$ must take the following form:
\begin{equation}\label{etasquareoddcccc}
\eta^2=\bigg{(}\begin{array}{ccc}
  0 & b \\
  -b & 0  \\
\end{array}\bigg{)},
\end{equation}
with $a=-b$. This can only be true if the matrix $\eta$ is of the form:
\begin{equation}\label{etasquareoddcccceta}
\eta=\bigg{(}\begin{array}{ccc}
  \sqrt{b} & -\sqrt{b} \\
  \sqrt{b} & \sqrt{b}  \\
\end{array}\bigg{)}.
\end{equation}
So in conclusion, we can add a trivial central charge to each of the $N_f!/(N_f-2)!+N_f$ different $N=2$ SUSY QM algebras. The question is whether there exists any other supersymmetric structure underlying this fermionic system equipped with non-trivial central charge and this will be the subject of the next section.

\section{Superconducting Cosmic Strings and Higher Extended SUSY structure}

Having found the $N_f!/(N_f-2)!+N_f$ different $N=2$ supersymmetric quantum mechanics algebras underlying the fermionic zero modes around the superconducting cosmic string, it is natural to ask whether there exists an enhanced supersymmetric structure underlying the system, or alternatively put, if these $N=2$ SUSY QM algebras can combine. As we shall demonstrate in detail in this section, the answer lies in the affirmative. As we shall see, depending on the number of fermion flavors, the supersymmetries we shall find are of the form of $N=4,6,8,...$, $d=1$ extended SUSY structures with non-trivial topological charges. The number ''$N$'' denotes the double of the total number of supercharges. In addition, as topological charges we will consider operators that are equal to the anticommutator of some Fermi charges (supercharges), following the terminology of references \cite{wittentplc,haag,fayet}. Moreover, when we have $N_f$ flavors for each fermion $(\psi_{\alpha}, \chi_{\beta} )$, we have an $N_s=2(\frac{N_f!}{(N_f-2)!}+N_f)$, $d=1$ extended SUSY structure with non-trivial topological charges. The topological charges we found do not commute with all the operators of the algebra unless some conditions are satisfied. This kind of topological charges often occurs in string theory frameworks, for example when a Wess-Zumino term is taken into account in a Ad$S_5\times S^5$ D-brane background with superalgebra $su(2,2|4)$ \cite{topologicalcharges2}. The latter is the super-isometry algebra of the Ad$S_5\times S^5$ background and the maximal extension is $osp(1|32)$. An interesting feature of all the cases we shall study is that the topological charges can be central charges if some conditions are satisfied. In the following we shall examine thoroughly some characteristic examples and also the general case.

\subsection{A Total Number of Three Fermions}

We start our study with a simple case first, in order to see all the features of the system in detail. We shall assume that we have one fermion $\psi_1$ and two fermions $\chi_i$, with $i=1,2$. Then, the corresponding supercharges that can be constructed are two, namely $\mathcal{Q}_{11}$ and $\mathcal{Q}_{12}$. The exact form of the supercharges can be obtained following the steps of section 2 and these are equal to:
\begin{equation}\label{wit2jdnhdsoft}
\mathcal{Q}_{ 1 1}=\bigg{(}\begin{array}{ccc}
  0 & D_{1 1} \\
  0 & 0  \\
\end{array}\bigg{)},{\,}{\,}{\,}
\mathcal{Q}_{ 1 2}=\bigg{(}\begin{array}{ccc}
  0 & D_{1 2} \\
  0 & 0  \\
\end{array}\bigg{)}
\end{equation}
with the operators $\mathcal{D}_{11}$ and $\mathcal{D}_{12}$ being equal to:
\begin{equation}\label{dmatrixnewsoft}
\mathcal{D}_{1 1}=\left(%
\begin{array}{cc}
  \partial_{1}+i\partial_{2} & -i\mathcal{M}^{\dag}_{1 1} \\
  i\mathcal{M}_{1 1} & \partial_{1}-i\partial_{2} \\
\end{array}%
\right),{\,}{\,}{\,}
\mathcal{D}_{1 2}=\left(%
\begin{array}{cc}
  \partial_{1}-i\partial_{2} & -i\mathcal{M}^{\dag}_{1 2} \\
  i\mathcal{M}_{1 2} & \partial_{1}+i\partial_{2} \\
\end{array}%
\right)
\end{equation}
The exact form of the Yukawa mass parameters $\mathcal{M}_{1 1}$ and $\mathcal{M}_{12}$ is of no importance for the moment. Recall that these parameters are the vacuum expectation values of the Higgs field and can be either constant ($\phi$-independent), or $\phi$-dependent. This issue and the outcomes of the $\phi$-dependence of the mass parameters will concern us in the end of this subsection. In order to reveal the underlying supersymmetric structure of this quantum system, we compute the following commutation and anticommutation relations and the results are:
\begin{align}\label{commutatorsanticomm}
&\{{{\mathcal{Q}}_{12}},{{\mathcal{Q}}_{12}}^{\dag}\}=2\mathcal{H}+\mathcal{Z}_{1212},
{\,}\{{{\mathcal{Q}}_{11}},{{\mathcal{Q}}_{11}}^{\dag}\}=2\mathcal{H}+\mathcal{Z}_{1111}
,{\,}\{{{\mathcal{Q}}_{11}},{{\mathcal{Q}}_{11}}\}=0,
\\ \notag &\{{{\mathcal{Q}}_{12}},{{\mathcal{Q}}_{12}}\}=0, {\,} \{{{\mathcal{Q}}_{11}},{{\mathcal{Q}}_{12}}^{\dag}\}=\mathcal{Z}_{1112},{\,}\{{{\mathcal{Q}}_{12}},{{\mathcal{Q}}_{11}}^{\dag}\}=\mathcal{Z}_{1211},{\,}\\ \notag
&\{{{\mathcal{Q}}_{12}}^{\dag},{{\mathcal{Q}}_{12}}^{\dag}\}=0,\{{{\mathcal{Q}}_{11}}^{\dag},{{\mathcal{Q}}_{11}}^{\dag}\}=0,{\,}\{{{\mathcal{Q}}_{11}}^{\dag},{{\mathcal{Q}}_{12}}^{\dag}\}=0,{\,}\{{{\mathcal{Q}}_{11}},{{\mathcal{Q}}_{12}}\}=0{\,}\\
\notag
&[{{\mathcal{Q}}_{12}},{{\mathcal{Q}}_{11}}]=0,[{{\mathcal{Q}}_{11}}^{\dag},{{\mathcal{Q}}_{12}}^{\dag}]=0,{\,}[{{\mathcal{Q}}_{12}},{{\mathcal{Q}}_{12}}]=0,{\,}[{{\mathcal{Q}}_{12}}^{\dag},{{\mathcal{Q}}_{12}}^{\dag}]=0
\end{align}
In the relation above, the operator $\mathcal{H}$ is equal to:
\begin{equation}\label{newsusymat1a}
\mathcal{H}=\left ( \begin{array}{ccccc}
  \partial_1^2+\partial_2^2 & 0 & 0 & 0\\
  0 & \partial_1^2+\partial_2^2 & 0 & 0 \\
  0 & 0 & \partial_1^2+\partial_2^2 & 0 \\
  0 & 0 & 0 & \partial_1^2+\partial_2^2 \\
\end{array}\right )
\end{equation}
while the operators $\mathcal{Z}_{1111}$ and $\mathcal{Z}_{1 2 1 2}$ are equal to:
\begin{equation}\label{newsusymat2a}
\mathcal{Z}_{ 1 1 1 1 }=\left ( \begin{array}{ccccc}
  \mathcal{Z}^1_{1 1 1 1 } & 0 \\
  0 & \mathcal{Z}^2_{1 1 1 1 }  \\
\end{array}\right ),{\,}{\,}{\,}\mathcal{Z}_{ 1 2 1 2 }=\left ( \begin{array}{ccccc}
  \mathcal{Z}^1_{1 2 1 2 } & 0 \\
  0 & \mathcal{Z}^2_{1 2 1 2 }  \\
\end{array}\right )
\end{equation}
with the operator $\mathcal{Z}^1_{1 1 1 1 }$ being equal to,
\begin{equation}\label{newsusymat3a}
\mathcal{Z}^1_{1 1 1 1 }=\left ( \begin{array}{cc}
 \mathcal{M}_{1 1}^{\dag}\mathcal{M}_{1 1} & -i(\partial_1+i\partial_2)\mathcal{M}_{1 1}^{\dag}-i\mathcal{M}_{1 1}^{\dag}(\partial_1+i\partial_2) \\
  i\mathcal{M}_{1 1}(\partial_1-i\partial_2)+i(\partial_1-i\partial_2)\mathcal{M}_{1 1} & \mathcal{M}_{1 1}\mathcal{M}_{1 1}^{\dag} \\
\end{array}\right )
\end{equation}
and also the operator $\mathcal{Z}^2_{1 1 1 1 }$ stands for,
\begin{equation}\label{newsusymat4a}
\mathcal{Z}^2_{1 1 1 1 }=\left ( \begin{array}{cc}
 \mathcal{M}_{1 1}^{\dag}\mathcal{M}_{1 1} & -i(\partial_1-i\partial_2)\mathcal{M}_{1 1}^{\dag}-i\mathcal{M}_{1 1}^{\dag}(\partial_1-i\partial_2) \\
  i\mathcal{M}_{1 1}(\partial_1+i\partial_2)+i(\partial_1+i\partial_2)\mathcal{M}_{1 1} & \mathcal{M}_{1 1}\mathcal{M}_{1 1}^{\dag} \\
\end{array}\right )
\end{equation}
The operators $\mathcal{Z}^1_{12 12}$ and $\mathcal{Z}^1_{12 12}$ can be obtained from those appearing in relations (\ref{newsusymat3a}) and (\ref{newsusymat4a}) by making the substitution $\mathcal{M}_{1 1} \rightarrow \mathcal{M}_{1 2}$. Finally, the operator $\mathcal{Z}_{1 1 1 2}$ is equal to:
\begin{equation}\label{newsusymat5a}
\mathcal{Z}_{1 1 1 2 }=\left ( \begin{array}{ccccc}
  \mathcal{Z}^1_{1 1 1 2 } & 0 \\
  0 & \mathcal{Z}^2_{1 1 1 2  }  \\
\end{array}\right )
\end{equation}
with the operator $\mathcal{Z}^1_{1 1 1 2 }$ being equal to:
\begin{equation}\label{newsusymat6a}
\mathcal{Z}^1_{1 1 1 2 }=\left ( \begin{array}{cc}
 \partial_1^2+\partial_2^2+\mathcal{M}_{1 1}^{\dag}\mathcal{M}_{ 1 2  } & -i(\partial_1+i\partial_2)\mathcal{M}_{1 2  }^{\dag}-i\mathcal{M}_{1 1}^{\dag}(\partial_1+i\partial_2) \\
  i\mathcal{M}_{1 1}(\partial_1-i\partial_2)+i(\partial_1-i\partial_2)\mathcal{M}_{1 2  } & \partial_1^2+\partial_2^2+ \mathcal{M}_{1 1}\mathcal{M}_{1 2 }^{\dag} \\
\end{array}\right )
\end{equation}
and in addition with $\mathcal{Z}^2_{1 1 1 2 }$ being equal to:
\begin{equation}\label{newsusymat7a}
\mathcal{Z}^2_{1 1 1 2 }=\left ( \begin{array}{cc}
 \partial_1^2+\partial_2^2+\mathcal{M}_{ i j }^{\dag}\mathcal{M}_{\alpha \beta} & -i(\partial_1-i\partial_2)\mathcal{M}_{\alpha \beta}^{\dag}-i\mathcal{M}_{ i j }^{\dag}(\partial_1-i\partial_2) \\
  i\mathcal{M}_{ i j }(\partial_1+i\partial_2)+i(\partial_1+i\partial_2)\mathcal{M}_{\alpha \beta} & \partial_1^2+\partial_2^2+\mathcal{M}_{ i j }\mathcal{M}_{\alpha \beta}^{\dag} \\
\end{array}\right )
\end{equation}
Notice that the operators $\mathcal{Z}_{1 1 1 1},\mathcal{Z}_{1 1 1 2}$ and $\mathcal{Z}_{1 2 1 2}$, are equal to anticommutators of Fermi charges (supercharges) and these are non-trivial topological charges \cite{wittentplc,fayet,topologicalcharges,topologicalcharges1}. 

The commutation and anticommutation relations computed in (\ref{commutatorsanticomm}) describe an $N=4$, $d=1$ superalgebra with non-trivial topological charges $\mathcal{Z}_{1 1 1 1},\mathcal{Z}_{1 1 1 2}$ an $\mathcal{Z}_{1 2 1 2}$. Hence, we can see that even in this simple case with three fermions, there exists a richer supersymmetric structure than the two $N=2$,$d=1$ of the system of three fermions. Now lets come to the issue of $\phi$-dependence of the Yukawa mass parameters. A rather more interesting case is when these parameters are constant and $\phi$-independent. This case can be realized if we consider the Yukawa couplings of the fermions at $r\rightarrow \infty$ and describes a phenomenologically realistic scenario \cite{supercondstrings}. In that case, the topological charges of the $N=4$ algebra, namely  $\mathcal{Z}_{1 1 1 1},\mathcal{Z}_{1 1 1 2}$ an $\mathcal{Z}_{1 2 1 2}$, commute with every element of the superalgebra, that is, with $\mathcal{H}$ and also with the supercharges $\mathcal{Q}_{11}$, $\mathcal{Q}_{12}$ and their conjugates. Therefore, the topological charges in that case are central charges of the SUSY QM algebra. We will discuss more thoroughly the issue of topological charges and central charges later on in this section. Let us now proceed to a more involved case, with a total number of four fermions. As we shall demonstrate, in this case the system has a much more rich supersymmetric structure.

\subsection{A Total Number of Four Fermions-Two Flavors for Each of the Fermions $(\psi_{\alpha},\chi_{\beta})$}

Following the line of research of the previous subsection, we assume in this case that we have two fermions $\psi_i$, $i=1,2$ and two $\chi_i$, $i=1,2$. In this case the total number of supercharges we can form increases drastically and therefore we have a much more rich supersymmetric structure with non-trivial topological charges. Let us see these, omitting the exact form of the supercharges for brevity and simplicity since these are too many. Nevertheless, in the next subsection we shall give general formulas of the supercharges and of the topological charge in order to have a complete picture of the supersymmetric structures.

The supercharges we can form in the case at hand are four supercharges, namely $\mathcal{Q}_{11},\mathcal{Q}_{12},\mathcal{Q}_{21},\mathcal{Q}_{22}$. Hence, we can form three kinds of different supersymmetric structures with non-trivial topological charges. Particularly, we can form $N=4$, $N=6$ and $N=8$ SUSY QM algebras with non-trivial topological charges. Let us see this in detail. Each $N=4$ SUSY QM algebra with topological charges is described by the following commutation and anticommutation relations:
\begin{align}\label{n4algbe1sjdjf}
&\{Q_{\alpha \beta},Q_{ij}^{\dag}\}=2\delta_{\alpha}^j\delta_i^{\beta}\mathcal{H}+Z_{\alpha \beta ij},{\,}{\,}{\,}{\,} \\ \notag &
\{Q_{\alpha \beta},Q_{i j}\}=0,{\,}{\,}\{Q_{\alpha \beta}^{\dag},Q_{i j}^{\dag}\}=0
\end{align}
and we take into account only two supercharges at a time. Therefore, from four supercharges we can form $\left ( \begin{array}{c}
 4 \\
  2 \\
\end{array}\right )=6$, $N=4$, $d=1$ supersymmetries with non-trivial topological charges, since the combination of two supercharges gives an $N=4$ SUSY algebra. In the same way, since the combination of three supercharges generates an $N=6$ supersymmetry, we have $\left ( \begin{array}{c}
 4 \\
  3 \\
\end{array}\right )=4$, $N=6$, $d=1$ superalgebras with non-trivial central charges. In the same fashion, taking all the four supercharges in account, we can form one $N=8$, $d=1$ SUSY QM algebra. We gather our results in table 1 where we can see the structure of each of the aforementioned supersymmetric structure.

\pagebreak

\begin{center} 
    \begin{tabular}{ | p{3cm} | p{2cm}  | p{6cm} |}
    \hline
    Supersymmetric Structure & Number of Different Supersymmetries & Groups of Supercharges that define the Algebra\\ \hline
    $N=4$ & 6 & $\begin{array}{c}
 (\mathcal{Q}_{11},\mathcal{Q}_{12}),(\mathcal{Q}_{21},\mathcal{Q}_{22}) \\
  (\mathcal{Q}_{11},\mathcal{Q}_{21}),(\mathcal{Q}_{12},\mathcal{Q}_{22}) \\
  (\mathcal{Q}_{11},\mathcal{Q}_{22}),
(\mathcal{Q}_{21},\mathcal{Q}_{12})\\
\end{array}$
     \\ \hline
    $N=6$ & 4 & $\begin{array}{c}
 (\mathcal{Q}_{11},\mathcal{Q}_{12},\mathcal{Q}_{21}),(\mathcal{Q}_{11},\mathcal{Q}_{12},\mathcal{Q}_{22})\\
  (\mathcal{Q}_{11},\mathcal{Q}_{21},\mathcal{Q}_{22}),(\mathcal{Q}_{21},\mathcal{Q}_{12},\mathcal{Q}_{22}) \\
\end{array}$
\\ \hline
    $N=8$ & 1 &  $(\mathcal{Q}_{11},\mathcal{Q}_{12},\mathcal{Q}_{21},\mathcal{Q}_{22})$ \\
    \hline
    \end{tabular}
    \\
    \bigskip 
    Table 1
\end{center}
In the case the Yukawa mass parameters are constant and $\phi$-independent, all the topological charges commute with every operator in the algebra. As in the case studied in the previous subsection, the topological charges in this case become central charges of the SUSY QM algebras.

\subsection{The General Case with $N_f$ Flavors for Each of the Fermions $(\psi_{\alpha},\chi_{\beta})$}

In the present case we assume that each of the fermions $(\psi_i,\chi_i)$ has $N_f$ flavors, that is $i=1,2,...N_f$. In such case we can form $N_s=N_f!/(N_f-2)!+N_f$ different supercharges of the following form:
\begin{equation}\label{wit2jdnhdgeneralc}
\mathcal{Q}_{\alpha \beta}=\bigg{(}\begin{array}{ccc}
  0 & D_{\alpha \beta} \\
  0 & 0  \\
\end{array}\bigg{)}
\end{equation}
with,
\begin{equation}\label{dmatrixnewgeneralcase}
\mathcal{D}_{\alpha \beta}=\left(%
\begin{array}{cc}
  \partial_{1}+i\partial_{2} & -iM^{\dag}_{\alpha \beta} \\
  iM_{\alpha \beta} & \partial_{1}-i\partial_{2} \\
\end{array}%
\right)_{2\times 2}
\end{equation}
and the values of the parameters are $\alpha ,\beta =1,2,...N_f$. Using these supercharges we can construct many N-extended SUSY QM algebras with many non-trivial topological charges and in addition $N=4,6,8,...$, depending on which supercharges we take into account. When we use all the $N_f$ flavors for each of the fermions $(\chi ,\psi )$, the largest algebra we can form is a single $2N_s$ SUSY QM algebra (recall $N_s=N_f!/(N_f-2)!+N_f$) with non-trivial topological charges, which we now present. The $N_s$ SUSY QM algebra is described by the following anticommutation relations:
\begin{align}\label{n4algbe1sjdjfgeneral}
&\{Q_{\alpha \beta},Q_{ij}^{\dag}\}=2\delta_{\alpha}^j\delta_i^{\beta}\mathcal{H}+Z_{\alpha \beta ij},{\,}{\,}{\,}{\,}\alpha ,\beta ,i ,j=1,2,..N_f \\ \notag &
\{Q_{\alpha \beta},Q_{i j}\}=0,{\,}{\,}\{Q_{\alpha \beta}^{\dag},Q_{i j}^{\dag}\}=0
\end{align}
where again in this general case, the operator $\mathcal{H}$ is equal to:
\begin{equation}\label{newsusymat12}
\mathcal{H}_{\alpha \beta \alpha \beta }=\left ( \begin{array}{ccccc}
  \partial_1^2+\partial_2^2 & 0 & 0 & 0\\
  0 & \partial_1^2+\partial_2^2 & 0 & 0 \\
  0 & 0 & \partial_1^2+\partial_2^2 & 0 \\
  0 & 0 & 0 & \partial_1^2+\partial_2^2 \\
\end{array}\right ).
\end{equation}
Notice that in relation (\ref{n4algbe1sjdjfgeneral}) appear many non-trivial topological charges which are equal to, or result from anticommutators of Fermi (supercharges) charges. These supercharges are non-trivial, as in all the previous studied cases and we shall present them in detail. The topological charges that result from anticommutators of a supercharge $\mathcal{Q}_{\alpha \beta }$ and its conjugate $\mathcal{Q}_{\alpha \beta }^{\dag}$ are of the form:
\begin{equation}\label{newsusymat22}
\mathcal{Z}_{\alpha \beta \alpha \beta }=\left ( \begin{array}{ccccc}
  \mathcal{Z}^1_{\alpha \beta \alpha \beta } & 0 \\
  0 & \mathcal{Z}^2_{\alpha \beta \alpha \beta }  \\
\end{array}\right ).
\end{equation}
with the operator $\mathcal{Z}^1_{\alpha \beta \alpha \beta }$ being equal to the following matrix:
\begin{equation}\label{newsusymat32}
\mathcal{Z}^1_{\alpha \beta \alpha \beta }=\left ( \begin{array}{cc}
 \mathcal{M}_{\alpha \beta}^{\dag}\mathcal{M}_{\alpha \beta} & -i(\partial_1+i\partial_2)\mathcal{M}_{\alpha \beta}^{\dag}-i\mathcal{M}_{\alpha \beta}^{\dag}(\partial_1+i\partial_2) \\
  i\mathcal{M}_{\alpha \beta}(\partial_1-i\partial_2)+i(\partial_1-i\partial_2)\mathcal{M}_{\alpha \beta} & \mathcal{M}_{\alpha \beta}\mathcal{M}_{\alpha \beta}^{\dag} \\
\end{array}\right )
\end{equation}
and correspondingly, the operator $\mathcal{Z}^2_{\alpha \beta \alpha \beta }$, is equal to the matrix:
\begin{equation}\label{newsusymat42}
\mathcal{Z}^2_{\alpha \beta \alpha \beta }=\left ( \begin{array}{cc}
 \mathcal{M}_{\alpha \beta}^{\dag}\mathcal{M}_{\alpha \beta} & -i(\partial_1-i\partial_2)\mathcal{M}_{\alpha \beta}^{\dag}-i\mathcal{M}_{\alpha \beta}^{\dag}(\partial_1-i\partial_2) \\
  i\mathcal{M}_{\alpha \beta}(\partial_1+i\partial_2)+i(\partial_1+i\partial_2)\mathcal{M}_{\alpha \beta} & \mathcal{M}_{\alpha \beta}\mathcal{M}_{\alpha \beta}^{\dag} \\
\end{array}\right )
\end{equation}
The topological charges that are directly equal to the anticommutator of a supercharge $\mathcal{Q}_{\alpha \beta }$ and the conjugate of another supercharge $\mathcal{Q}_{i j}^{\dag}$, are equal to:
\begin{equation}\label{newsusymat52}
\mathcal{Z}_{\alpha \beta i j }=\left ( \begin{array}{ccccc}
  \mathcal{Z}^1_{\alpha \beta i j } & 0 \\
  0 & \mathcal{Z}^2_{\alpha \beta i j }  \\
\end{array}\right ).
\end{equation}
with $\mathcal{Z}^1_{\alpha \beta i j }$ being the matrix:
\begin{equation}\label{newsusymat62}
\mathcal{Z}^1_{\alpha \beta i j }=\left ( \begin{array}{cc}
 \partial_1^2+\partial_2^2+\mathcal{M}_{\alpha \beta}^{\dag}\mathcal{M}_{ i j } & -i(\partial_1+i\partial_2)\mathcal{M}_{ i j }^{\dag}-i\mathcal{M}_{\alpha \beta}^{\dag}(\partial_1+i\partial_2) \\
  i\mathcal{M}_{\alpha \beta}(\partial_1-i\partial_2)+i(\partial_1-i\partial_2)\mathcal{M}_{ i j } & \partial_1^2+\partial_2^2+ \mathcal{M}_{\alpha \beta}\mathcal{M}_{ i j }^{\dag} \\
\end{array}\right )
\end{equation}
and additionally the operator $\mathcal{Z}^2_{\alpha \beta i j }$ is equal to:
\begin{equation}\label{newsusymat72}
\mathcal{Z}^2_{\alpha \beta i j }=\left ( \begin{array}{cc}
 \partial_1^2+\partial_2^2+\mathcal{M}_{ i j }^{\dag}\mathcal{M}_{\alpha \beta} & -i(\partial_1-i\partial_2)\mathcal{M}_{\alpha \beta}^{\dag}-i\mathcal{M}_{ i j }^{\dag}(\partial_1-i\partial_2) \\
  i\mathcal{M}_{ i j }(\partial_1+i\partial_2)+i(\partial_1+i\partial_2)\mathcal{M}_{\alpha \beta} & \partial_1^2+\partial_2^2+\mathcal{M}_{ i j }\mathcal{M}_{\alpha \beta}^{\dag} \\
\end{array}\right )
\end{equation}
The exact value of the Yukawa mass parameters (Higgs vacuum expectation values) are not so significant for the present analysis of the form of the topological charges. However, if these are constant and $\phi$-independent, the topological charges commute with every operator of the $2N_s$, $d=1$ SUSY QM algebra and consequently they become central charges of the algebra. Having $N_s$ supercharges at our disposal, we can form many N-extended SUSY QM algebras with $N\leq N_s$. In table 2 we provide the exact number of supersymmetries given the number $N$.
\begin{center} 
    \begin{tabular}{ | p{3cm} | p{3cm}  | }
    \hline
    Supersymmetric Structure & Number of Different Supersymmetries \\ \hline
    $N=4$ & $\left ( \begin{array}{c}
 N_s \\
  2 \\
\end{array}\right )$ 
     \\ \hline
    $N=6$ &$\left ( \begin{array}{c}
 N_s \\
  3 \\
\end{array}\right )$  
\\ \hline
    $N=8$ & $\left ( \begin{array}{c}
 N_s \\
  4 \\
\end{array}\right )$  \\
    \hline
    \vdots & \vdots
     \\ \hline
     $N=2(N_s-1)$ & $\left ( \begin{array}{c}
 N_s \\
  N_s-1 \\
\end{array}\right )$  
     \\ \hline
     $N=2N_s$ & 1 
     \\ \hline
    
    \end{tabular}
    \\
    \bigskip 
    Table 2
\end{center}
with $\left ( \begin{array}{c}
 \nu \\
 \kappa \\
\end{array}\right )$ given by the very well known formula:
\begin{equation}\label{syndyasmopi}
\left ( \begin{array}{c}
 \nu \\
 \kappa \\
\end{array}\right )=\frac{\nu!}{(\nu-\kappa)!\kappa !}
\end{equation}
Hence, the underlying supersymmetric structures of the fermionic zero modes system around a superconducting string are quite many, depending on the number of the fermion flavors. An issue that deserves our attention is the existence of non-trivial topological charges that become central charges for constant and $\phi$-independent Yukawa mass parameters. We shall briefly discuss this issue in the next subsection.

\subsection{A Brief Discussion on Topological Charges and Central Charges}

Let us briefly recapitulate what we have at hand so far. The system of fermionic zero modes around a superconducting cosmic string has a rich underlying supersymmetric structure which, depending on the total number of flavors, can vary from many $N=4$ to a single $2N_s$ (with $N_s=N_f!/(N_f-2)!+N_f$) one dimensional supersymmetric algebra with non-trivial topological charges. Interestingly enough, when the Yukawa mass parameters (which are Higgs vacuum expectation values) are constant and $\phi$-independent, then the topological charges commute with every operator of the superalgebra and therefore become central charges of the algebra.  

The existence of topological charges in supersymmetric algebras was noticed in \cite{wittentplc} were the terminology topological charge was first used. As noted by the author of reference \cite{fayet}, the topological charges cannot be considered as central charges, since these do not commute with all the operators of the superalgebra. Moreover, as was also noted in \cite{fayet}, the topological charges are symmetries of the field theory but not of the $S$-matrix of the theory. This seems to be the case that applies to our findings. These topological charges we found in the general case are not symmetries of the $S$-matrix, but indicate some additional external symmetry of the field theory that describes the zero modes around the defect. It is not accidental that the theoretical framework used in \cite{wittentplc}, was a supersymmetric algebra in the presence of topological defects and the presence of topological charges in such frameworks are unavoidable. This interplay between supersymmetry and topological defects is particularly interesting and was also pointed out in \cite{topologicalcharges}. In addition, as pointed out in reference \cite{topologicalcharges1}, the existence of a ''central'' charge that does not commute with the rest operators of the algebra indicates a possible non-linear richer supersymmetric structure. It would be interesting to try to find if such a structure exists in our case too, but we defer this task to a future work.

Before closing let us comment that non-commuting topological charges occur in string theory contexts, for example in Ad$S_5\times S^5$ D-brane background with superalgebra $su(2,2|4)$ \cite{topologicalcharges2}. The topological charges in these frameworks have also non trivial commutation relations with the elements of the superalgebra.

\section*{Concluding Remarks}

In this article the focus was on the system of zero modes around a superconducting cosmic string and the relation of this quantum system to supersymmetry. Particularly, we revealed a hidden supersymmetric structure that underlies every system of fermions that consists of two different fermions ($\psi, \chi$), taking into account their interaction with the Higgs field through Yukawa couplings. As we demonstrated, having $N_f$ flavors for each fermion participating in the system, we can construct $N_s=N_f!/(N_f-2)!+N_f$ one dimensional $N=2$ supersymmetries with zero central charge. In addition, these supersymmetries are unbroken, something that is guaranteed from the existence of zero modes, or equivalently from the string superconductivity. Moreover, we studied the Witten index of the supersymmetric system and we investigated its behavior under compact perturbations caused by a background magnetic field. As we saw, the index is invariant if the magnetic field contains only one component and particularly if it is of the form $\mathcal{B}=(0,0,\mathcal{B}_z)$. In addition, we answered the question if the string is superconducting in the case $I_q=0$, and as we saw the string is superconducting in the case the $N=2$, $d=1$ supersymmetry is unbroken. We also presented some higher order reducible representations of the $N=2$ SUSY QM algebra, using the $N_s=N_f!/(N_f-2)!+N_f$ supercharges and also we extended each $N=2$ algebra to an $N=2$, $d=1$ algebra with a trivial central charge.

Another issue we addressed in this article is whether there exists any other supersymmetric structure apart from the $N_s=N_f!/(N_f-2)!+N_f$ one dimensional supersymmetries. As we demonstrated, the answer lies in the affirmative. Particularly, depending on the fermion flavor number $N_f$ (or depending on the total number of fermions $2N_f$), there exist multiple higher $N$ one dimensional supersymmetries with non-trivial topological charges, with $N=4,6,8,...,2N_s$. In the most general case, the results are gathered in table 2 of the previous section. When the Yukawa couplings mass parameters (Higgs vacuum expectation values) are constant and $\phi$-independent, the topological charges become central charges of the theory, since these commute with every other operator of the algebra. In addition, we discussed the existence of non-trivial topological charge in a supersymmetric algebra and its possible implications.

Although the initial system had no supersymmetric structure, it seems that there exists a rich non-trivial supersymmetric underlying structure that gets even more involved when the number of flavors increases. This behavior has been pointed out in the literature, see for example \cite{oiko1}. Particularly, it seem that global supersymmetry plays no obvious role in the process we studied in this article. Our approach revealed an underlying symmetry of the field theory, a symmetry that is not a symmetry of the S-matrix of the theory.

Let us point out here that in general, spacetime supersymmetry in $d>1$ dimensions and
SUSY QM, which is an $d=1$ supersymmetry, are in principle different concepts, with the only possible correlation being the fact that extended (with $N = 4, 6...$) SUSY QM models are obtained by the dimensional reduction of $N = 2$ and $N = 1$ Super-Yang Mills models down to one dimension  \cite{ivanov}. Nevertheless, the complex supercharges of $N = 2$, $d=1$ SUSY QM are not related in any way to the generators of spacetime supersymmetry and therefore, SUSY QM does not directly relate fermions and bosons, with fermions and bosons considered as representations of the super-Poincare graded Lie algebra in four dimensions. On the other hand, the SUSY QM supercharges render the Hilbert space of quantum states a $Z_2$ graded vector space, in the simplest case of $N=2$ SUSY QM. Furthermore, these supercharges generate the transformations between the Witten parity eigenstates. In our case, the one dimensional supersymmetry is not a spacetime symmetry nor a symmetry of the S-matrix.

Extended supersymmetries are quite important and are met quite frequently in the study of Super-Yang-Mills systems. Particularly, the $N=4$ supersymmetric algebra is frequently met in string theory contexts, since extended (with $N=4,6...$) supersymmetric quantum mechanical models result by the dimensional reduction $N=2$ and $N=1$ Super-Yang Mills theories to one temporal dimension. Moreover, extended supersymmetries are directly related to super-extensions of integrable models (Calogero-Moser systems), and in addition to super-extensions of Landau-type models \cite{ivanov}.  

In principle, a complex problem may be simplified if we find techniques to reveal all possible internal symmetries. In addition, if instead of symmetries we discover some regularity or some repeating underlying pattern, we accomplish an even more deep insight to the problem. As was pointed out in reference \cite{rossi}, the existence of the fermionic zero modes in the presence of a topological string (vortex) defect can be related to the presence of a hidden underlying symmetry of the fermionic system at hand. The material presented in \cite{rossi} is different in spirit from problems of localized fermions on higher dimensional defects like branes (for an important stream of papers on this, see \cite{liu} and references therein). The symmetry found in reference \cite{rossi}, was some kind of supersymmetry \cite{rossi}. Therefore, it is quite intriguing the fact that the fermionic zero modes around a cosmic string are directly associated to $N$-extended one dimensional supersymmetries, with non-trivial supercharges. These SUSY QM algebras can be associated to some even more complex algebraic symmetry of the field theory. Lastly, in relation to the higher algebraic structure, let us comment that the presence of non-trivial topological charges in the presence of a defect (cosmic string), deserves a more detailed study on how these topological charges appear. Their existence indicates some richer, probably non-linear, algebraic structure related to supersymmetry, yet to bet found. We hope we can address this issue in the near future.

\end{document}